\pdfoutput=1
\documentclass{article}
\usepackage{spconf,amsmath,graphicx}
\usepackage{pgfplots}
\pgfplotsset{compat=newest}
\usetikzlibrary{plotmarks}
\usetikzlibrary{arrows.meta}
\usepgfplotslibrary{patchplots}

\newcommand{\iec}{i.\,e.,\ }

\newcommand{\Figure}[1]{Fig.~\ref{#1}}
\newcommand{\Table}[1]{Table~\ref{#1}}

\def\imgs2{5.2 cm}
\title{A Data-driven Cognitive Salience Model for Objective Perceptual Audio Quality Assessment}
\name{Pablo M. Delgado $^{\star}$, Jürgen Herre $^{\dagger \star}$ }

\address{$^{\star}$ International Audio Laboratories Erlangen $^{\ddagger}$, Am Wolfsmantel 33, 91058 Erlangen, Germany \protect\thanks{$^{\ddagger}$ A joint institution of the Friedrich-Alexander Universität Erlangen-Nürnberg (FAU) and Fraunhofer IIS, Germany.} \\
$^{\dagger}$ Fraunhofer IIS, Am Wolfsmantel 33, 91058 Erlangen, Germany \\
Correspondence should be addressed to pablo.delgado@audiolabs-erlangen.de\\
}
\begin{document}
%
\maketitle
\begin{abstract}
Objective audio quality measurement systems often use perceptual models to predict the subjective quality scores of processed signals, as reported in listening tests. Most systems map different metrics of perceived degradation into a single quality score predicting subjective quality. This requires a quality mapping stage that is informed by real listening test data using statistical learning (\iec a data-driven approach) with distortion metrics as input features. However, the amount of reliable training data is limited in practice, and usually not sufficient for a comprehensive training of large learning models. Models of cognitive effects in objective systems can, however, improve the learning model. Specifically, considering the salience of certain distortion types, they provide additional features to the mapping stage that improve the learning process, especially for limited amounts of training data. We propose a novel data-driven salience model that informs the quality mapping stage by explicitly estimating the cognitive/degradation metric interactions using a salience measure. Systems incorporating the novel salience model are shown to outperform equivalent systems that only use statistical learning to combine cognitive and degradation metrics, as well as other well-known measurement systems, for a representative validation dataset.
\end{abstract}
\section{Introduction}
\label{sec:intro}

Objective audio quality measurement systems are tools for evaluating or monitoring the perceived quality of processed music or speech signals by means of signal analysis and feature extraction. These systems are expected to save time and resources by predicting results of subjective listening tests, which are considered the standard method for estimating quality.

The majority of the current objective quality systems incorporate models of human auditory perception, mostly peripheral effects like auditory masking, motivated by their successful use in the field of audio coding \cite{RixTaslp}. Quality degradation is a multidimensional phenomenon, and it can be driven by multiple factors such as perceived roughness, noisiness and linear distortions ---among others--- in the processed signals \cite{erne2001perceptual, dick2017generation}. In many popular subjective quality assessment methods, quality degradation in multiple dimensions is rated using a single mean opinion score (MOS) \cite{MUSHRA,ITUP800}. Likewise, objective perceptual quality systems map different distortion measures (DM) derived from the perceptual model into one single quality score by means of multivariate regression, using subjective MOS as training data. The combination of different DM into an objective quality score is considered to be a basic data-driven (and task-dependent) model of cognition \cite{ThiedePHD}. The relative importance of each of the factors influencing perceived quality degradation is related to distortion \textit{salience}, and is driven by cognitive effects \cite{ThiedePHD,bregman1994auditory}. Salience influences the perceived severity of an artifact in an audio signal and is usually implicitly modeled by the mapping stage, although it has also been explicitly modeled \cite{vincent2012improved}.

Cognitive models have been considered in objective perceptual quality measurement systems  to model the salience of a DM \cite{beerends1996the,Par2056_2019}. The cognitive model outputs (here, cognitive effect metrics - CEM) can interact with DM by scaling their measured intensity. Systems with cognitive-corrected DM have shown inconsistent performance \cite{ThiedePHD}, since salience rarely depends on a single cognitive effect \cite{sporer1997objective}.
Ideally, performing multivariate statistical learning with the DM and CEM \cite{barbedo2005a} as direct inputs can lead to a more complete model of salience. The learning models may find meaningful CEM/DM interactions describing overall quality given enough training data available. However, training data is usually costly and scarce \cite{DelgadoPEAQ}, and learning models may not have enough data to perform DM-to-quality-score mapping and determine meaningful CEM/DM interactions at the same time, causing the model to overfit \cite{ThiedePHD}.

Based on the assumption that particular cognitive effects influence distortion salience, we propose a novel cognitive salience model (CSM) as an extension to the multivariate statistical learning approach. While also data-driven, the CSM separates the quality mapping from the CEM/DM interaction model. The CEM/DM interactions are informed and optimized using a salience cost function, and not the cost function used by the multivariate regression mapping. The training data in this stage only informs a small number of CEM/DM interaction model parameters. In this way, the CSM model is expected to outperform multivariate learning models that use DM and CEM as direct inputs for the training dataset sizes typically available in practice.

\section{Method}
\label{sec:method}

\begin{figure}[htb]
  \makebox[0.8\textwidth][l]{
      \includegraphics[width=0.48\textwidth]{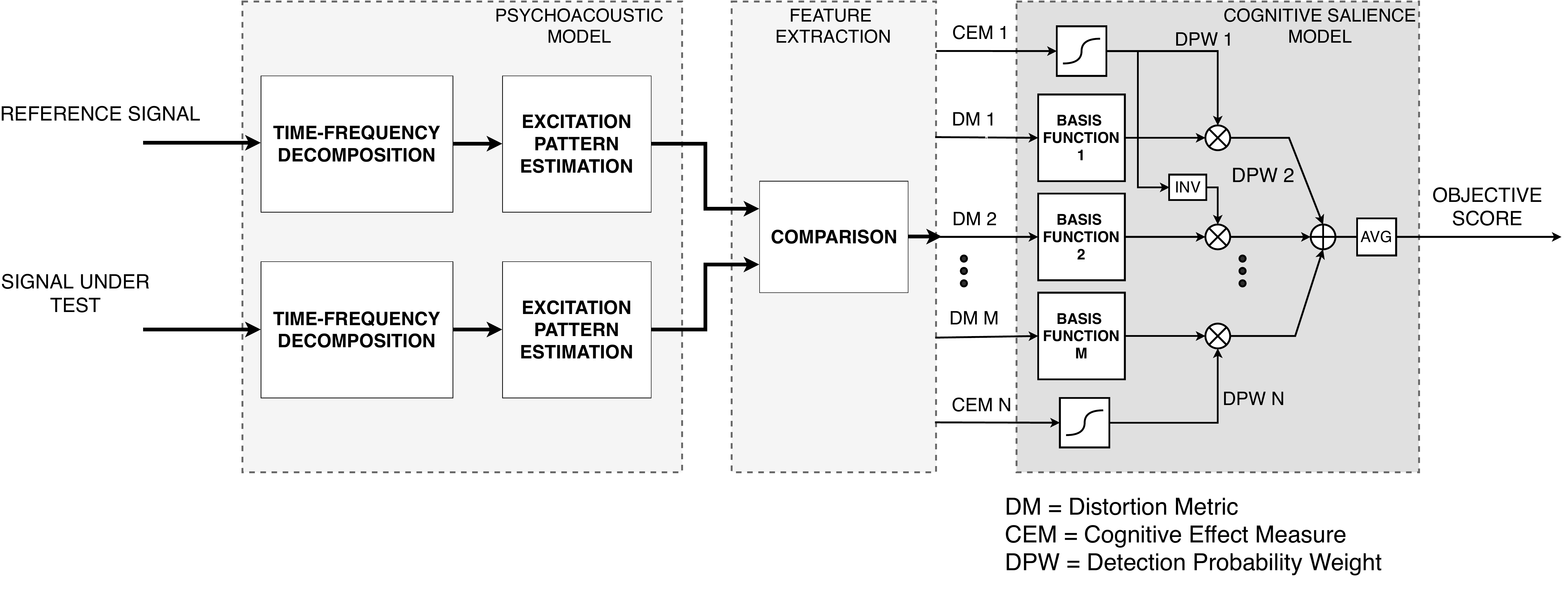}
  }
\caption{Block diagram of the proposed objective perceptual quality measurement system.}
\label{fig:block_diagram}
\end{figure}

As basis for our contribution, we propose a \textit{full-reference} \cite{RixTaslp} objective quality measurement system (\Figure{fig:block_diagram}) composed by a perceptual model (PM) which produces a series of perceptually-motivated DM and CEM, and a cognitive salience model  that maps said metrics into a single objective quality score. The objective quality score is expected to be a predictor of an overall MOS, as graded by listening test subjects.

\subsection{Perceptual Model}

The perceptual model used corresponds to our own implementation \cite{DelgadoPEAQ} of the one used by the Perceptual Evaluation Audio Quality (PEAQ) method \cite{PEAQ}, in its \textit{advanced} version. The perceptual model compares reference (REF) and signal under test (SUT) in the transformed psychophysical domain (time-pitch-loudness or modulation). Multiple comparisons using peripheral and cognitive models in the psychophysical domain produce the different DM.

\begin{table}[t]
  \centering
    \resizebox{\columnwidth}{!}{%
      \begin{tabular}{l|c|c}
        \textbf{DM} & \textbf{Associated MOV} &\textbf{Distortion Type} \\ \hline
        LinDist & AvgLinDist\_A & Linear distortions  \\
        ModDiff & RmsModDiff\_A & Modulation disturbances \\
        NoiseLoudness & RmsNoiseLoudAsym\_A & Added noise in SUT \\
        MissingComponents & RmsMissingComponents\_A & Missing components in SUT \\
        EHS  & EHS\_B & Harmonic structure of the error \\
        SegNMR  & Segmental NMR\_B & Noise-to-mask ratio \\
      \end{tabular}
    }
				\caption{Distortion metrics used and their related PEAQ MOVs (Tables 4 and 17 in \cite{PEAQ}).}
	\label{tab:distmetrics}
\end{table}

The distortion metrics produced by the perceptual model in PEAQ can be averaged over time and frequency to produce a single-number estimate per comparison, termed the Model Output Values (MOV). We perform frequency averaging and time averaging over 2-second segments on the distortion metrics, leaving the overall time averaging (AVG) over the whole length of the audio excerpt to a later stage in the CSM. The distortion metrics we used are based on corresponding MOVs, and are shown in \Table{tab:distmetrics}.

\begin{table}[t]
  \centering
    \resizebox{\columnwidth}{!}{%
      \begin{tabular}{l|c}
        \textbf{CEM} & \textbf{Description} \\ \hline
        EPN & Perceptual streaming measure from \cite{barbedo2005a}, Eq. (30)  \\
        PDEV & Informational masking from \cite{barbedo2005a}, Eq. (31) \\
        probSpeech & Probability of speech-like signal from \cite{SpeechVsMusicQuality} \\
      \end{tabular}
    }
				\caption{Cognitive effect metrics used in this work.}
	\label{tab:cognimetrics}
\end{table}

In addition to DM, CEM (\Table{tab:cognimetrics}) can be derived from the perceptual model. We consider a model for two important cognitive effects from the Auditory Stream Analysis field \cite{bregman1994auditory}: \textit{perceptual streaming} (PS) and \textit{informational masking} (IM). PS indicates whether some disturbance in the SUT is perceived as a separate auditory event from the masker signal itself (\iec REF). In this case, the disturbance is likely to be perceived as more annoying than when it forms one single percept with the masker signal. In contrast, IM increases the masking threshold when the signal and disturbance variations in time and frequency are large enough, due to an increased rate of information reaching the auditory cortex. PS is thought to weaken the effect of IM and they can be seen as competing effects \cite{beerends1996the}. We implement the measures of PS and IM as proposed in \cite{barbedo2005a}. In addition, speech and music quality perception are believed to make use of some slightly different processing mechanisms in the auditory nerve \cite{SpeechVsMusicQuality}. Therefore, the speech-music classifier proposed in \cite{USACSpeechClas} will be used as CEM. The classifier analyzes the signal in the early stages of the perceptual model, before the psychophysical representation transformations.

\subsection{Cognitive Salience Model}
\label{ref:CSM}
The proposed CSM combines the different DM into a single objective score via a weighted sum based on the outputs of different CEM. The weighting mechanism is expected to model distortion salience in quality perception. 

\label{sec:BF}
In order to perform the weighted sum of DM into a single output score, the individual DM stemming from the perceptual model need to be mapped to a target subjective quality scale though a basis function (BF). The BFs were estimated using Multivariate Adaptive Regression Splines (MARS) \cite{Jekabsons_areslab}, mapping the associated MOVs in \Table{tab:distmetrics} to MOS (using the MUSHRA procedure \cite{MUSHRA}) in the database described in \cite{dick2017generation}, produced with isolated audio coding artifacts. The isolated audio coding artifacts minimize distortion type interaction, favoring the independence of each estimated BF.


As with many psychoacoustic results, cognitive effects modeled after controlled laboratory experiments with simple signals are not likely to completely generalize to overall quality perception in real use case signals \cite{ThiedePHD}. In real use cases, distortion interactions and experimental conditions in listening tests are most likely to mask the influence of small cognitive effect sizes in perceived quality, decreasing the \textit{probability} that subjects detect this effect. Likewise, perceived audio quality might no longer be influenced by larger measured effect sizes. For example, the IM effect has been reported to saturate with larger masker information amounts \cite{lutfiAndOh}.

We use psychometric functions \cite{kingdom2016psychophysics} ---in particular, the logistic function family--- to relate the CEM outputs and the actual weights to model the described phenomena. The CEM transformed by the psychometric functions are termed the \textit{detection probability weights} (DPW). Since a cognitive effect size can actually hinder the salience of a distortion type, an inverse operation (INV) in the detection probability domain can be defined as $1-DPW$ to address these cases.

\subsection{Optimization of Detection Probability Parameters}
\label{sec:optimization}

The interaction model parameters between CEM and MOVs will be estimated using a data-driven approach using listening test data. In our model, the interaction model parameters are specifically two parameters of the involved logistic functions: curve steepness and crossover midpoint. A third parameter, the curve's maximum value, will be therefore fixed at 1, emulating a detection probability. We carry out an exhaustive search optimization procedure for the two parameters in each of the DPW relating a CEM with a DM. The optimization cost function will be based on a salience measure. 

The salience $\mathcal{S}$ for a DM $m,\mbox{ }m=1\ldots M$ in a signal $j, \mbox{ } j=1\ldots J$ is defined as the Pearson correlation coefficient between the respective DM basis function $BF_m$ and the MOS $y$ for all available audio treatments $i=1\ldots I$:

\begin{equation}
  \resizebox{0.9\columnwidth}{!}{
  $\mathcal{S}_m(j)
    = \frac{\sum_{i=1}^{I} ( y_{ij}-\overline{y}_{j} )( BF_{mij}-\overline{BF}_{mj} )}
  {\sqrt{\sum_{i=1}^{I} (y_{ij}-\overline{y}_{j})^2} \sqrt{\sum_{i=1}^{I} (BF_{mij}-\overline{BF}_{mj})^2}}$
  }.
\end{equation}
The salience metric is calculated for averaged values over time for the duration of the signals. The stronger the BF/MOS correlation is for a given DM in a signal, the more it is assumed that the DM describes quality degradation, and therefore the higher its salience.

The optimization cost function describes the covariance of the salience metric for a DM against the DPW output for said DM across all signals in the database. It is a metric of DM/CEM interaction. The optimization of the DPW parameters is expected to improve the interaction model. The correlation coefficient is defined as:

\begin{equation}
  \resizebox{0.9\columnwidth}{!}{
  $\mathcal{C}_m
    = \bigg|\frac{\sum_{j=1}^{J} ( S_m(j)-\overline{S}_{m} )( DPW_{m}(j)-\overline{DPW}_{m} )}
  {\sqrt{\sum_{j=1}^{J} (S_{m}(j)-\overline{S}_{m})^2} \sqrt{\sum_{j=1}^{J} (DPW_{m}(j)-\overline{DPW}_{m})^2}}\bigg|$
  }.
\end{equation}
The stronger the correlation, the better the DPW will predict a given DM salience over the signals in the database.

\section{Experiment Design}
\label{sec:Experiment}

We train and validate the proposed system with an extensive audio coding quality database of speech, music, and mixed signals at a sampling rate of 48 kHz under different treatments at bitrates ranging from 16kbps to 96kbps in stereo mode \cite{USACdatabase}. A total of 168 quality scores for 24 signals and 7 treatments for training, and a total of 216 pooled scores of 24 signals and 9 treatments (different than those of the training database) were used for validation. More than 25000 individual MUSHRA scores between the two used tests were used for the pooling.

We carry out the optimization procedure for the DPW on each CEM/BF interaction described in Section \ref{sec:optimization} using the Verification Test 2. We keep only the $DPW_m$ in the CSM that provided strong values of $\mathcal{C}_m$ after optimization, or CEM/BF interactions that have been established in psychoacoustic research. The system is then validated on the Verification Test 1 database along with some other variants.

\subsection{Evaluated System Variants}
\label{sec:proposed}

The main purpose of the proposed CSM system is to establish an interaction model between cognitive effects and distortion salience that will improve objective measurement performance. We propose three variants to account for these interactions. The first one will consider both the CEM and the DM as input variables of a mapping stage (emulating the approach used in \cite{barbedo2005a}) trained with the database used in Section \ref{sec:BF} and the training database of Section \ref{sec:Experiment}. The mapping stage consists of an Artificial Neural Network (ANN) with the same settings as listed for the advanced version of PEAQ \cite{RevisionBS1387}. The resulting model will be labeled $DM + CEM$. The second and the third variants correspond to the same implementation of the CSM of Section \ref{ref:CSM} but differing in the use of DPW for the DM/CEM interactions. The $PROPOSED$ system weights the BF directly by the CEM. The $PROPOSED\mbox{ }(Opt.)$ system weights the BF using the optimized DPW from Section \ref{sec:optimization}. The performance difference should indicate if there is any gain in modeling detection probability for cognitive effects. Additionally, the performance on the same validation data will be reported for two objective quality measurement systems: ViSQOL Audio \cite{Visqol_soft} and PEAQ Advanced DI (distortion index) \cite{RevisionBS1387}. PEAQ's DI output encompasses a broader quality range than its ODG output, the latter being quality-scale dependent \cite{ThiedePHD,DelgadoPEAQ}.

Overall performance is rated in terms of MOS/Objective score correlation $R$, and prediction error $RMSE^*$ as recommended in \cite{EvalObjective}, including techniques meant to reduce possible biases and quality gradients between the evaluated systems' training and validation databases.

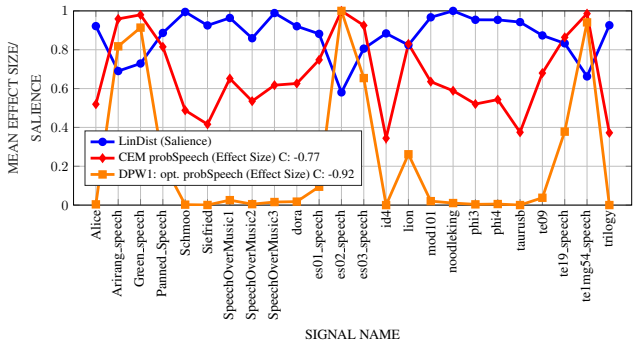
\begin{figure}[htb]
  \makebox[\textwidth][l]{
    \resizebox{0.53\width}{!}{
%
%
\begin{tikzpicture}

\begin{axis}[%
width=14cm,
height=4.875cm,
at={(0cm,0cm)},
scale only axis,
xmin=0,
xmax=25,
xtick={1,2,3,4,5,6,7,8,9,10,11,12,13,14,15,16,17,18,19,20,21,22,23,24},
xticklabels={{Alice},{Arirang\_speech},{Green\_speech},{Panned\_Speech},{Schmoo},{Siefried},{SpeechOverMusic1},{SpeechOverMusic2},{SpeechOverMusic3},{dora},{es01\_speech},{es02\_speech},{es03\_speech},{id4},{lion},{mod101},{noodleking},{phi3},{phi4},{taurusb},{te09},{te19\_speech},{te1mg54\_speech},{trilogy}},
xticklabel style={rotate=90},
xlabel style={font=\normalsize},
xlabel={SIGNAL NAME},
ymin=0,
ymax=1,
ylabel style={font=\color{white!15!black}, align=center},
ylabel={MEAN EFFECT SIZE/\\[1ex]SALIENCE },
axis background/.style={fill=white},
xmajorgrids,
ymajorgrids,
legend style={at={(0.02,0.1)},anchor=south west,legend cell align=left, align=left, draw=white!15!black},
legend style={font=\small},
]
\addplot [color=blue, line width=2.0pt, mark=otimes*, mark options={solid, blue}]
  table[row sep=crcr]{%
1	0.921689302794165\\
2	0.690160528153531\\
3	0.729118520088729\\
4	0.886560187174597\\
5	0.994806672119575\\
6	0.925168209407104\\
7	0.963829766644456\\
8	0.859210810397561\\
9	0.988942751866792\\
10	0.920970097431821\\
11	0.881739276965217\\
12	0.580492226661912\\
13	0.805825143543113\\
14	0.88405334788852\\
15	0.824079981189414\\
16	0.966951726037525\\
17	1\\
18	0.954168647734673\\
19	0.953912559558891\\
20	0.942357624626653\\
21	0.87367200006372\\
22	0.83288991848072\\
23	0.662591544624924\\
24	0.926104265700772\\
};
\addlegendentry{$\text{LinDist} \text{ (Salience) }$}

\addplot [color=red, line width=2.0pt, mark=diamond*, mark options={solid, red}]
  table[row sep=crcr]{%
1	0.519395131092052\\
2	0.958839894090783\\
3	0.979663764717109\\
4	0.814959876840239\\
5	0.488100864019267\\
6	0.416183762332928\\
7	0.651583451017625\\
8	0.535197609970899\\
9	0.617504375565135\\
10	0.626242420612211\\
11	0.747848879016157\\
12	1\\
13	0.925263261034612\\
14	0.344208429137637\\
15	0.829953743813112\\
16	0.635674114173775\\
17	0.588729858126584\\
18	0.520380099534952\\
19	0.543266163908604\\
20	0.374970945867713\\
21	0.679891498709342\\
22	0.863845670062008\\
23	0.985800618470931\\
24	0.372320376492107\\
};
\addlegendentry{CEM probSpeech (Effect Size) C: -0.77}

\addlegendentry{DPW1: opt. probSpeech (Effect Size) C: -0.92}

\addplot [color=orange, line width=2.0pt, mark=square*, mark options={solid, orange}]
  table[row sep=crcr]{%
1	0.00375051999439402\\
2	0.817700672454315\\
3	0.913936606593578\\
4	0.219773054924863\\
5	0.00229448721181233\\
6	0.000613508418611306\\
7	0.0254966711340117\\
8	0.00476605811702463\\
9	0.0157783771041201\\
10	0.0178570106276827\\
11	0.094531537780277\\
12	1\\
13	0.654039695778106\\
14	0\\
15	0.26170208117847\\
16	0.0203977369088841\\
17	0.0104576822082636\\
18	0.00380748586915404\\
19	0.00537756556111127\\
20	0.000189730935207951\\
21	0.0377719553722926\\
22	0.378214198051791\\
23	0.940851212929987\\
24	0.000169970816173378\\
};
\addlegendentry{$\text{probSpeech}_\text{d}\text{pw (Effect Size)}$}

\addlegendentry{R: -0.92747}

\end{axis}
\end{tikzpicture}%
    }
  }
\caption{Linear Distortion Salience, Mean Effect Size for a CEM (probSpeech) of the optimization database signals and associated DPW for the CEM/DM interaction.}
\label{fig:plot_CEM_vs_MOV}
\end{figure}

\section{RESULTS AND DISCUSSION}
\label{sec:results}

\begin{table*}[t!]
  \centering
    \resizebox{1.6\columnwidth}{!}{%
      \small
      \begin{tabular}{l|c|c|c|c}
        \textbf{Weight} & \textbf{CEM} &\textbf{Target DM} & \textbf{C (CEM/DPW)}& \textbf{Equation} \\ \hline
        DPW1 & probSpeech & LinDist            & -0.77 / -0.92 & DPW1 = 1- probSpeech\_th\_lin \\
        DPW2 & probSpeech & NoiseLoudness      & 0.67 / 0.80   & DPW2 = probSpeech\_th\_nl \\
        DPW3 & probSpeech & MissingComponents  & -0.20 / -0.37   & DPW3 = 1-DPW2 \\
        DPW4 & EPN           & LinDist       & -0.40 / -0.70 & DPW4 = 1-EPN\_th\_lin \\
        DPW5 & EPN           & SegmentalNMR       & 0.1 / 0.25 & DPW5 = (EPN\_th\_sgm)(1-PDEV\_th\_sgm) \\
        DPW5 & PDEV          & SegmentalNMR       & -0.18 / -0.21 & - \\
      \end{tabular}
    }
				\caption{Selected interactions and cost function values (CEM and DPW) for the proposed CSM.}
	\label{tab:resultsOptim}
\end{table*}

The interaction between the salience measure of the $LinDist$ DM, the $probSpeech$ CEM and its optimized DPW for the optimization database signals is shown in \Figure{fig:plot_CEM_vs_MOV}. A strong CEM/Salience covariance $(\mathcal{C}=-0.77)$ supports the assumption that CEM can predict a given DM importance in quality degradation. Linear distortions are more salient in music than in speech signals for the analyzed database. The optimized psychometric function improved the cognitive effect and distortion interaction $(\mathcal{C}=-0.92)$.

The results of the optimization procedure and DPW choice for the proposed models can be seen in Table \ref{tab:resultsOptim}. The $\_th$ suffix indicates the ``thresholded" CEM with the optimized logistic function considering the target DM salience. The remaining target DM that have not been listed are considered to contribute in equal parts to the final objective measure, with no DPW weighing.

Some conclusions consistent with previous reports can be drawn from Table \ref{tab:resultsOptim}. Considering DPW2 and DPW3, added noise is more salient for speech signals. The importance of added noise measures for speech has been confirmed by the objective speech quality literature \cite{PESQ}. Likewise, missing components are considered less annoying in this context. Coefficient DPW4 predicts that linear distortions are less important when perceptual streaming is likely taking place. In general, linear distortions are considered less annoying than added noise \cite{ThiedePHD}. More so if PS exacerbates the perception of added noise. Regarding DPW5, the implementation of PS/IM weights as competing effects (as in \cite{beerends1996the}) on an instantaneous measure of the noise-to-mask ratio improved performance of the model despite weak individual $\mathcal{C}$ values. Future work might contemplate how to better capture these higher order interactions between CEM and DM using data analysis.

\begin{table}[t!]
  \centering
      \begin{tabular}{l|c|c}
        \textbf{System} & $\mathbf{R}$ & $\mathbf{RMSE^*}$ \\ \hline
        ViSQOL NSIM & 0.82 & 5.6  \\
        PEAQ DI & 0.69 & 8.1  \\
        DM + CEM & 0.84 & 5.1 \\
        PROPOSED  & 0.86 & 4.6 \\
        PROPOSED (Opt.) & \textbf{0.90} & \textbf{3.7} \\
      \end{tabular}
				\caption{System validation performance metrics \cite{EvalObjective} for the systems proposed in Section \ref{sec:proposed}. Best performance metrics in bold.}
	\label{tab:resultsTotal}
\end{table}

\Table{tab:resultsTotal} shows the validation results for the proposed systems of Section \ref{sec:proposed}. The proposed CSM variant with the DPW outperforms all investigated systems for the data used. Note that the pure combination of CEM and DPW into an ANN ($DM+CEM$) did not perform as well as the proposed CSM. The ANN learning algorithm did not seem to find interactions as meaningful as is our perceptually-motivated and data-driven model.

\section{Conclusion and Future Work}

The overall inclusion of the presented cognitive effect metrics in objective perceptual quality measurement has proven to be promising. Systems including cognitive effects showed an improved performance with respect to baseline systems, encouraging the use of such metrics. The proposed CSM and the data-driven optimization of CEM/DM interactions seem to be a reasonable approach for incorporating cognitive metrics with the dataset sizes available in practice, outperforming the use of more traditional statistical learning techniques that consider CEM and DM as input features.

Future work should contemplate further validation in larger and more diverse databases, especially for use cases other than perceptual audio coding. Additionally, the inclusion of other cognitive effect models, interactions with other distortion metrics and using alternative salience measures may be further investigated.


\clearpage
\bibliographystyle{IEEEbib}
\bibliography{paper_references}

\end{document}